\documentclass[%
 aip,
 amsmath,amssymb,
reprint,
groupedaddress
]{revtex4-1}

\draft 

\usepackage{graphicx}
\usepackage{dcolumn}
\usepackage{bm}

\usepackage[utf8]{inputenc}
\usepackage[T1]{fontenc}
\usepackage{mathptmx}

\begin{document}


\title{High-Power, Fiber-Laser-Based Source for Magic-Wavelength Trapping in Neutral-Atom Optical Clocks}



\author{William J. Eckner}
\email[]{william.eckner@colorado.edu}

\author{Aaron W. Young}

\author{Nathan Schine}

\author{Adam M. Kaufman}

\affiliation{JILA, University of Colorado and National Institute of Standards and Technology,
and Department of Physics, University of Colorado, Boulder, Colorado 80309, USA}


\date{\today}

\begin{abstract}
We present a continuous-wave, 810 nm laser with watt-level powers. Our system is based on difference-frequency generation of 532 nm and 1550 nm fiber lasers in a single pass through periodically poled lithium niobate (PPLN). We measure the broadband spectral noise and residual intensity noise to be compatible with off-resonant dipole trapping of ultracold atoms. Given the large bandwidth of the fiber amplifiers, the output can be optimized for a range of wavelengths, including the strontium clock-magic wavelength of 813 nm. Furthermore, with the exploration of more appropriate nonlinear crystals, we believe there is a path toward scaling this proof-of-principle design to many watts of power, and that this approach could provide a robust, rack-mountable trapping-laser for future use in strontium-based optical clocks.

\end{abstract}


\maketitle

\section{Introduction}

Alkaline-earth atoms are a cornerstone of neutral-atom optical clocks,\cite{mcgrew2018atomic} as well as studies of degenerate quantum gases,\cite{stellmer2009bose} diatomic molecules,\cite{mcguyer2015precise} and cavity QED.\cite{norcia2016superradiance} In comparison to alkali atoms, a distinguishing benefit of alkaline-earths is their ultranarrow `clock transition.' To take full advantage of the opportunities that this transition affords, experiments must confine atoms in so-called `magic wavelength' (MWL) traps, for which the light shifts on the ground and clock-excited states are approximately equal, and perturbations of trapping light on the clock-transition frequency are minimized.\cite{takamoto2005optical}

In this work, we consider applications to strontium clocks, for which the MWL is near 813 nm. Given the atomic polarizability of strontium at this wavelength, experiments often require a few watts of optical power in order to load and image large atomic samples. However, high optical powers cannot come at the cost of a trapping laser's intensity noise or spectral purity. For example, power fluctuations at even multiples of the trap frequency can parametrically heat atoms, causing Doppler broadening and loss, and optical power that is far-detuned from the MWL can cause light shifts on the clock frequency. As such, incoherent features in a laser's optical spectrum -- such as those caused by amplified spontaneous emission (ASE) -- must be compatible with state-of-the-art optical clocks, whose fractional-frequency uncertainties have been characterized at or below the $10^{-18}$ level.\cite{bloom2014optical}

In strontium clocks, Ti:Sapphire lasers have become a standard solution for meeting these challenges, as their high-power output has manageable ASE,\cite{fasano2021characterization} and can support exceptional atomic lifetimes.\cite{hutson2019engineering, young2020half} However, Ti:Sapphire lasers can be more challenging to operate outside of low-vibration, laboratory environments. Alternatively, recent experiments, especially those aimed at studying portable strontium clocks, have relied on the direct output of low-power diode lasers (often with a build-up cavity around the atoms),\cite{bowden2019pyramid, schiller2012space, akatsuka2010three} or tapered amplifiers (TAs). While TAs have a small footprint, they are limited to comparatively lower powers than those attainable with Ti:Sapphire lasers.\cite{schkolnik2020direct, poli2014transportable} 

\begin{figure}
\includegraphics[width=8.5cm]{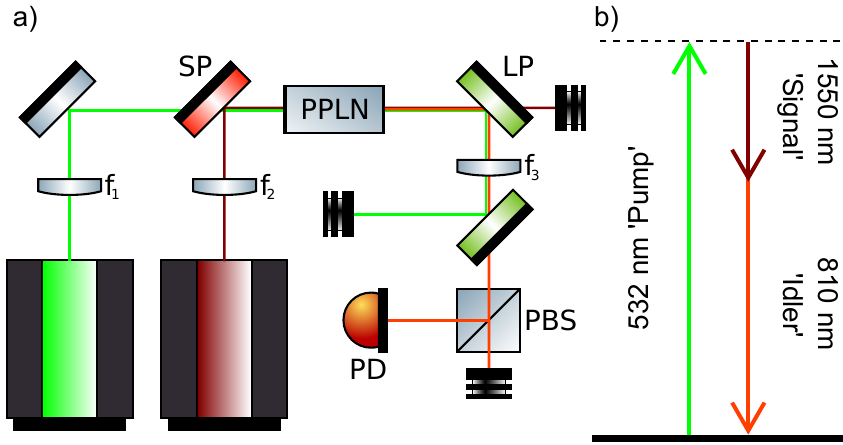}
\caption{\label{fig:Optics} \textbf{a)} Simplified schematic of our optical layout. Red optics labeled `SP' (green, `LP') refer to short-pass (long-pass) dichroic mirrors. All measurements of generated idler light are made after a polarizing beam splitter (`PBS'). Power measurements are made with a thermal power meter,\footnote{S425C from Thorlabs} and RIN measurements are made with a fixed-gain photodetector (labeled `PD'\footnote{PDA10A2 from Thorlabs}). \textbf{b)} Laser wavelengths for the input (pump, signal) and generated (idler) lasers. As demanded by energy conservation, DFG yields two signal photons, and one idler photon for each pair of pump and signal photons that is converted.}
\end{figure}

Here, we develop a laser system that can address these performance considerations by using a commercially available PPLN crystal\footnote{MSFG1-0.5-40, from Covesion} to generate light at the difference frequency between two high-power, continuous-wave (CW) fiber-laser sources. The 532 nm `pump' laser for this difference-frequency generation (DFG) process\cite{boyd2020nonlinear} can provide up to 10 W of power, with a linewidth $<$200 kHz.\footnote{ALS-GR-532-10-I-CP-SF, from Azurlight Systems} For our `signal' light, we use an erbium-doped fiber amplifier (EDFA), \footnote{Koheras BOOSTIK, K592-000-018, from NKT} seeded by a single-frequency OEM laser \footnote{Koheras BASIK MIKRO, K022- 40-1565.50-FCPGT-0,5-SSPC-0, from NKT} with a center wavelength of about 1550 nm. This laser provides 15 W of power, with a Lorentzian linewidth $<$200 Hz. We then convert these into `idler' light near 810 nm. The specific wavelengths we use in this work were chosen due to their short-term availability within our lab, but one can attain similar EDFAs within a few tens of nanometers around 1550 nm, enabling production of MWL lasers for strontium optical clocks. 

With this scheme, we reach optical powers up to 1.3 W near 810 nm, and we find that the generated light exhibits an ASE background that is suppressed by $>40$ dBc/Hz near the center wavelength, as measured after a single-mode optical fiber. If the laser were centered on the 813 nm MWL, we estimate that the resulting frequency shifts on the strontium clock transition could be reduced below the $10^{-18}$ level, for typical trap depths. We also measure relative intensity noise (RIN) that is $<-120$ dBc/Hz for frequencies between 30 Hz and 100 kHz, which is within the bandwidth of acousto-optic modulators used to actively suppress RIN, often by $>20$dB. Based on manufacturer specifications, we believe that the RIN above 100 kHz is $<-150$ dBc/Hz. 

We also note that the measured limitations on power scaling and ASE can be traced to our choice of nonlinear crystal and fiber lasers, respectively. Since both higher-damage-threshold crystals\cite{kurimura2005qpm,tovstonog2008thermal} and lower-noise fiber lasers are available, the performance measured here -- while already suitable for use in metrology -- is not inherent to our DFG scheme, and could be improved in future realizations. 

\section{Setup}

Single-pass, quasi-phase-matched frequency conversion with periodically poled crystals is a powerful technology for a range of demanding applications in atomic physics,\cite{oates2007stable, wilson2011750, hankin2014two, bai2019autler, bridge2016tunable} as it can imbue high-power visible laser light with the narrow linewidths, and robust operation of longer-wavelength fiber lasers. The viability of this approach for a desired output wavelength relies primarily on the availability of appropriate source lasers,\cite{fu2017review} and nonlinear crystals that can be precisely poled for a given quasi-phase-matching process.\cite{xu2012quasi} In this work, we take advantage of the fact that the frequency of 813 nm, MWL light is close to the difference-frequency of 532 nm and 1550 nm lasers, which have each been the subject of extensive study, and can be obtained at high power, as well as with impressive spectral purity.\cite{samanta2007high, delavaux1995multi} 

\begin{figure}
\includegraphics[width=8.5cm]{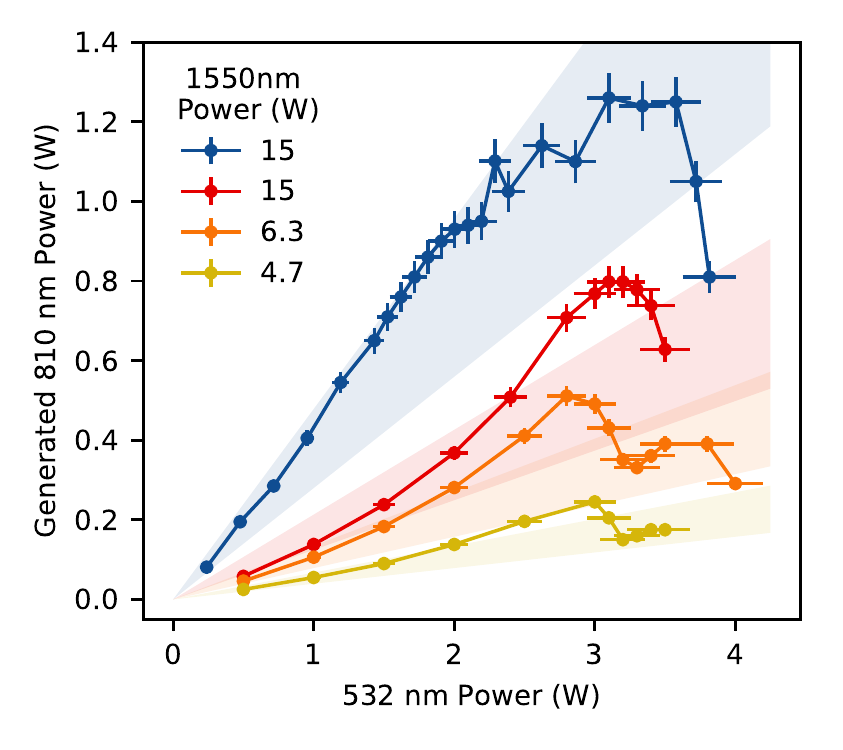}
\caption{\label{fig:Pow} Idler power (810 nm) versus pump power (532 nm) at a few different signal powers (1550 nm). In anticipation of crystal-heating at high intensities, we test this DFG process for two different pairs of beam waists, so as to probe the onset of thermal effects. The red, orange, and yellow curves are collected with larger pump and signal waists of $(w_p, w_s) = (82.9(3) \ \mu \text{m}, \ 147(2) \ \mu \text{m})$, as measured in free-space ($f_1 = 500$ mm, $f_2 = 300$ mm). The blue curve represents conversion with waists that are twice as small ($f_1 = 250$ mm, $f_2 = 150$ mm), and closer to optimal conditions. Error bars on data come from 5\% uncertainty in measurements with our thermal power meter. Boyd-Kleinman theory predicts idler powers in the shaded regions, where the uncertainty is dominated by limited knowledge of the parameter $d_{33}$, as discussed in the main text.}
\end{figure}

Fig.~\ref{fig:Optics} shows the optics diagram for our design (Fig.~\ref{fig:Optics}a), as well as a schematic depiction of the DFG process (Fig.~\ref{fig:Optics}b), which outputs 1$\times$idler photon (810 nm) and 2$\times$signal photons (1550 nm) for each pair of pump (532 nm) and signal photons that is converted. The expected performance can be estimated with Boyd-Kleinman theory, \cite{boyd1968parametric} and in the simplified case of normal beam incidence on the nonlinear crystal, the amount of converted power should be given by \cite{broyer1985intracavity, canarelli1992continuous, goldberg1995difference}
\begin{align}
P_1 = \frac{4 l \omega_1^2 k_2 k_3 d_{\text{eff}}^2}{\pi \epsilon_0 (k_2 + k_3) n_1 n_2 n_3 c^3} h(\mu, \xi) P_2 P_3 , \label{eq:eff}
\end{align}
where $P_i, \omega_i, n_i$ are the power, angular frequency, and index of refraction in the crystal, and the labels $i=1,2,3$ correspond to the idler, signal, and pump, respectively. Additionally, $c$ is the speed of light, $k_i = \omega_i / c$, $l$ is the length of the nonlinear crystal, $\epsilon_0$ is the vacuum permittivity, and $h(\mu, \xi)$ is the Boyd-Kleinman focusing function (see Fig.~\ref{fig:h_parameter} in the Appendix). The arguments of $h$ are the focusing parameter $\xi = l/b$ (where the confocal parameter $b$ is defined as twice the Rayleigh range of the beam) and $\mu = k_2/k_3$. We estimate the index of refraction at each wavelength using the extended Sellmeier coefficients in \textit{Lin et al}.\cite{lin2017extended} Finally, $d_{\text{eff}}$ is the effective nonlinear coefficient for our phase-matching process, and is related to the $d_{33}$ coefficient of our chosen crystal -- PPLN, with $ \text{length} = 40 \ \text{mm}, \text{width} =  0.5 \ \text{mm}, \  \text{height} = 0.5 \ \text{mm}$ -- as well as the duty cycle of the poling.\cite{hu2013engineered} Typically, $d_{\text{eff}} \approx 15 \ \text{pm/V}$ for PPLN at visible and near-IR wavelengths, but has been quoted over a range of values, sometimes as high as $17 \ \text{pm/V}$. \cite{goldberg1995difference, shoji1997absolute, miller199742, sowade2010nonlinear} In order to capture the uncertainty in the value $d_{\text{eff}}$, which depends on the manufacturing process for the crystal, as well as the wavelength dependence of $d_{33}$, we present Boyd-Kleinman theory curves in Fig.~\ref{fig:Pow} as shaded regions, bounded above and below by curves that assume values of $17 \ \text{pm/V}$ and $13 \ \text{pm/V}$, respectively. 

Eq.~\ref{eq:eff} is valid when the confocal parameters of the pump ($b_3$) and signal ($b_2$) are equal. However, in this work we test two different pairs of beam sizes, and in both cases $b_3/b_2 = 0.93$. For the larger (smaller) pair of pump and signal beam waists, this leads to a 6\% (2\%) difference between $h(\xi_3, \mu)$ and $h(\xi_2, \mu)$. To account for this, calculations in Fig.~\ref{fig:Pow} take the average confocal parameter $b = (b_2 + b_3)/2$. 

We observe reasonable agreement between the expected (Eq. \ref{eq:eff}) and generated idler power, shown in Fig.~\ref{fig:Pow}, when the pump power is less than about 2.5 W. However, at high pump powers, there is significant deviation, and ultimately loss of conversion. We believe that this is due to absorption and heating in the PPLN crystal. One indication of thermal effects comes from the change in the optimum setpoint temperature on our PPLN oven controller (see Fig.~\ref{fig:Temps} in the Appendix), which we reoptimize before each datapoint in Fig.~\ref{fig:Pow}. Additionally, the deviations in idler power lead, at first, to higher than expected output powers, but only for the case when the beam waist is larger than optimal to begin with. This is consistent with an intermediate regime where the self-focusing of the beam in the crystal actually increases the value of $h(\mu, \xi)$. Potential sources for significant heating are direct absorption of 532 nm pump light,\cite{schwesyg2011optical} as well as green-induced absorption of the infrared signal or idler light.\cite{moore2002self}

As a result of these thermal effects, we observe power instability over the course of tens of minutes when operating at the largest power of 1.3(1) W. To avoid these issues, the measurements presented in the rest of this report operate at 1 W of 810 nm power (see blue curve in Fig.~\ref{fig:Pow}), where the output power was stable over many hours, and optimal alignment of the pump and signal did not drift from day to day. 

\section{Relative Intensity Noise}

\begin{figure}
\includegraphics[width=8.5cm]{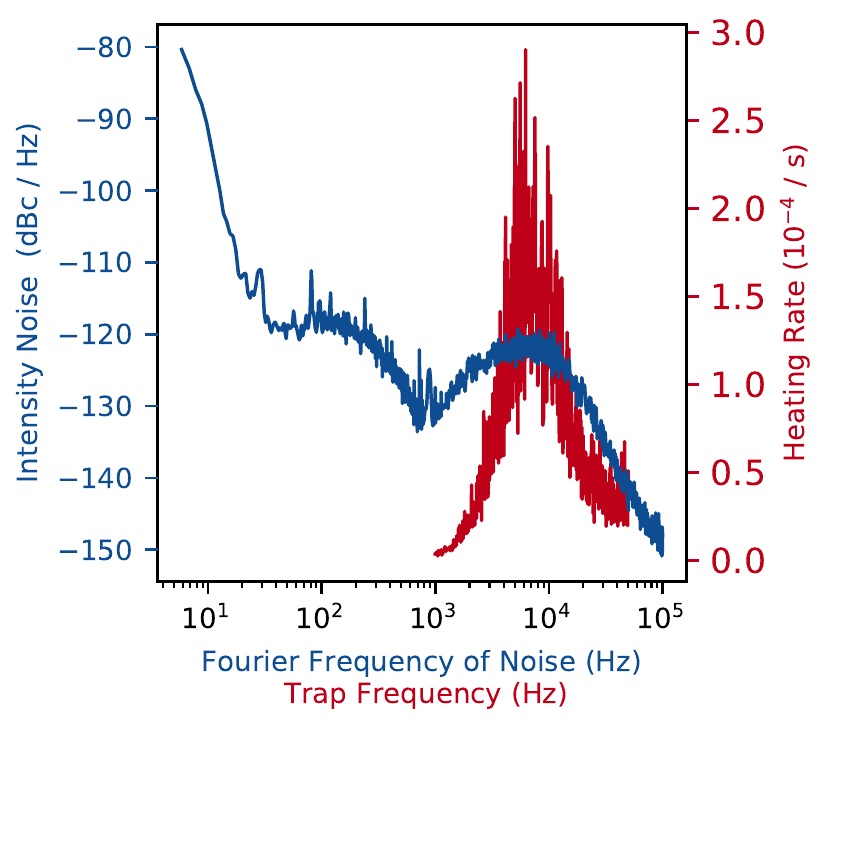}
\caption{\label{fig:RIN} \textbf{Left axis} (blue) relative intensity noise (RIN). \textbf{Right axis} (red) calculated exponential heating rates for atoms confined in harmonic potentials whose depth fluctuates according to the measured RIN, and as a function of the oscillator trap frequency. We note that the RIN presented here is for a free-running system, and the noise could be significantly reduced with active stabilization.\cite{wang2020reduction}}
\end{figure}

One of the most important characteristics for a trapping laser in ultracold-atom experiments is its level of RIN. In particular, amplitude noise at twice the trap frequency can cause parametric heating, which is sometimes the primary limitation on trapping lifetimes. In this section we determine the intensity noise on the generated light, shown in Fig.~\ref{fig:RIN}. Measurements at frequencies above 100 kHz were consistent with the noise floor of our spectrum analyzer and, based on manufacturer specifications, should be $<$ -150 dBc/Hz (see Fig.~\ref{fig:RIN_Appendix} in the Appendix).

The trap frequencies for optical lattice and tweezer clocks are typically in the range of 10 to 100 kHz, and as such low RIN at similar frequency scales is critical for minimizing the effects of parametric heating. The RIN profile we observe in Fig.~\ref{fig:RIN} is in good agreement with the specifications for our fiber-laser sources (Fig.~\ref{fig:RIN_Appendix} in the Appendix). Additionally, the noise on this free-running output is concentrated within the bandwidth of most acousto-optic modulators, which could provide an additional >20 dB of suppression. Using the Fokker-Planck model in \textit{Gehm et al.} one can calculate the expected heating rate for traps made with our system's output,\cite{gehm1998dynamics} and find a worst-case heating rate of roughly $2.9 \times 10^{-4}$ per second, for 6 kHz trap frequencies, as shown in Fig. 3. By numerically solving the rate equations associated with parametric heating in a harmonic trap,\cite{blatt2015low} we find that this corresponds to a three-dimensional motional ground state lifetime on the order of a few minutes. As this timescale is much longer than those associated with other technical sources of heating, we believe that this laser should be compatible with the observed trap lifetimes of $>100$ s in optical lattices or tweezers, such as those utilized in our lab.\cite{young2020half, hutson2019engineering}

\section{Optical Spectrum}

\begin{figure}[h!]
\includegraphics[width=8.5cm]{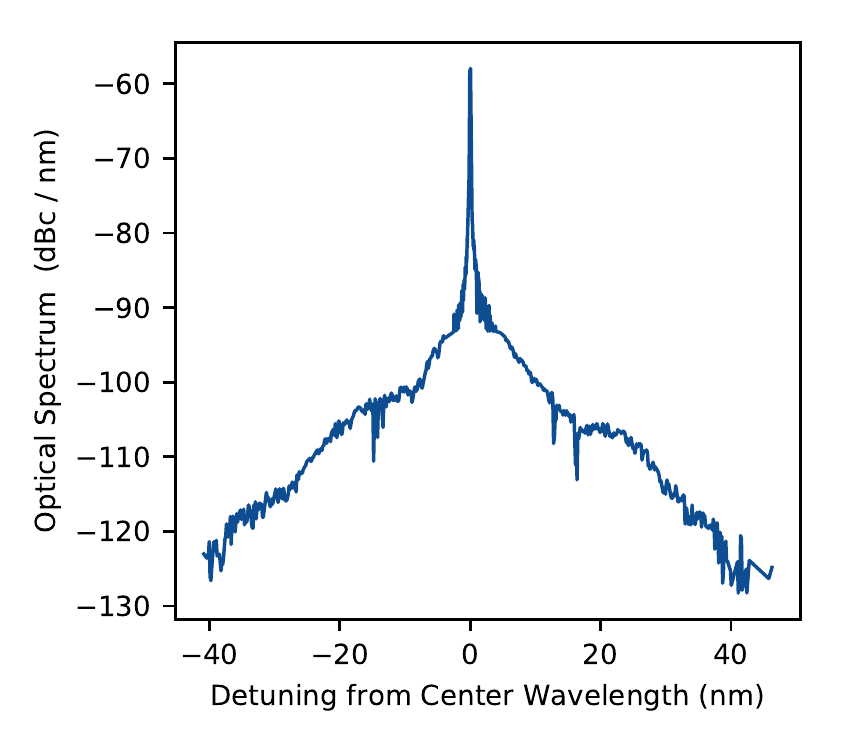}
\caption{\label{fig:OSA} Optical-power spectrum read on an OSA, after coupling into a polarization-maintaining, single-mode fiber.\footnote{P5-630PM-FC-1 from Thorlabs} The curve is normalized to the peak value, and wavelengths are referenced to the center of the spectrum. The minimum resolution of our measurement is 0.05 nm, which is much greater than the linewidth of the laser. This incommensurately coarse analyzer resolution biases our initial measurement, giving a lower peak value than would a finer probe. To compensate for this, we do not plot bins that are within 0.05nm of the center, and assume that the power measured within them is entirely contained in a 200 kHz carrier (an upper bound on our estimate of the laser's linewidth based on manufacturer specifications). We then normalize the spectrum to the inferred density of this narrow central feature.}
\end{figure}

Along with RIN, the spectral purity of trapping lasers can be of central importance, particularly for sensitive atomic-clock experiments. To understand the output of our system, we measure the generated 810 nm idler on an optical spectrum analyzer (OSA). The results are shown in Fig.~\ref{fig:OSA}. A salient characteristic of fiber-laser spectra for use in ultracold-atom experiments is the presence of a broad ASE pedestal.\cite{belotelov2020lattice} The relevant metric for evaluating the effects of ASE in an atomic clock is the fractional-frequency shift of the clock transition for a given laser intensity, assuming that the spectrum is centered on the MWL. For example, it was determined that the ASE from a typical TA operating at the MWL induced a fractional-frequency shift of roughly $4 \times 10^{-19}$ on the strontium clock transition, per $E_r$ of trap depth.\cite{ bilicki2017strontium} Here $E_r$ is the recoil energy of an 813 nm trap photon. As an estimate on the effects of ASE, we calculate the polarizability of strontium's ground and clock states over the range of wavelengths in Fig.~\ref{fig:OSA}, and determine the resulting fractional shift on the clock transition to be on the order of $10^{-19}/E_r$, when considering wavelengths further than 0.05 nm from the carrier. This effect is similar to calculations for TA spectra after passing through a single-mode fiber. It is important to consider the fact that single-mode fibers can provide substantial ASE filtering,\cite{belotelov2020lattice, bilicki2017strontium} and can also introduce alignment-dependent asymmetries in the optical spectrum. For future work with a permanent DFG system integrated into our experimental apparatus, it will be important to measure the optical spectrum before and after it is filtered by a single-mode fiber.

In order to assess the compatibility with metrological applications, we note that the most accurate strontium clocks currently reach trap depths of around $45 \ E_r$, and uncertainties at the $10^{-18}$ level.\cite{bothwell2019jila} Given our laser's estimated ASE-induced shift of $10^{-19}/E_r$, we hypothesize that it could be used in clock experiments with state-of-the-art accuracy. However, given the imprecision of our OSA measurement, especially near the center frequency, future investigations of this claim will utilize an optical heterodyne measurement with another narrow laser,\cite{fasano2021characterization} as well as direct observation of light shifts using our lab's optical-tweezer clock.\cite{young2020half}

\section{Discussion}

We have demonstrated a high-power, low-noise 810 nm laser, based on frequency conversion of rack-mountable fiber lasers. With the specification of lower noise pump and signal lasers, we believe this DFG process could synthesize light at the 813 nm MWL for strontium's clock transition, and would be compatible with fractional-frequency shifts that are suppressed below the $10^{-18}$ level, and trapping lifetimes $>$100 s. Additionally, it will be possible to scale this system to higher optical powers and lower levels of ASE with the use of higher-damage threshold nonlinear crystals\cite{kurimura2005qpm,tovstonog2008thermal} and lower noise fiber amplifiers, both of which are commercially available. 

The simplicity of our scheme makes it a particularly appealing option for current designs of strontium clocks that will operate in space,\cite{schkolnik2020direct} or in studies of geodesy, for which spacious, low-noise conditions are not feasible.\cite{grotti2018geodesy} Another strength of this DFG-based design is that the 810 nm light should inherit a linewidth that is the quadrature sum of the pump and signal, which can each be exceptionally narrow for fiber lasers. These linewidths would then allow for minimal heating in optical lattices, which are formed by standing-wave interference patters, and are sensitive to laser phase noise. Finally, we hypothesize that a similar solution should be possible in ytterbium lattice clocks, where the necessary infrared, `signal' wavelength would fall within the bandwidth of low-noise, high-power thulium amplifiers. 

\begin{acknowledgments}
We acknowledge helpful discussions with A.C. Wilson and J. Ye as well as close readings of the manuscript by W.F. McGrew and C.J. Kennedy. We also thank R. B. Hutson
of the Ye group for sharing his polarizability calculations. Any mention of commercial products is for information only; it does not imply recommendation or endorsement by NIST. This material is based upon work supported by NIST, AFOSR, and ARO. W.J.E. and N.S. acknowledge support from NDSEG and the NRC research associateship program, respectively.
\end{acknowledgments}

\appendix 
\section{Focusing Parameter $\bm{h(\mu, \xi)}$}

\begin{figure}[h!]
\includegraphics[width=8.5cm]{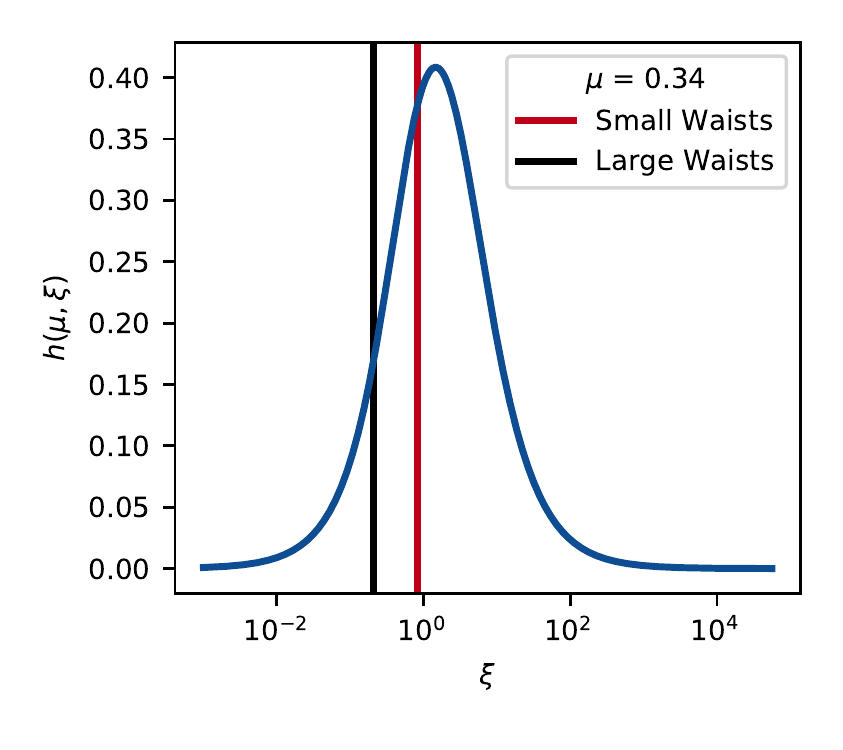}
\caption{\label{fig:h_parameter} The Boyd-Kleinman focusing function $h(\mu, \xi)$ for a range of focusing parameters $\xi$. Vertical lines correspond to the focusing parameters used in this work. }
\end{figure}

The focusing parameter $h(\mu, \xi)$ in Eq~\ref{eq:eff} is given by 
\begin{widetext}
\begin{equation}
h(\mu, \xi) = \frac{1}{2\xi} \int_0^\xi \text{d}\tau'' \int_{-\xi}^{\xi} \text{d} \tau' \frac{1 + \tau' \tau''}{(1 + \tau' \tau'')^2 + \frac{1}{4} (1 - \mu)/(1 + \mu) + [(1 + \mu)/(1 - \mu)]^2 (\tau' - \tau'')^2}
\end{equation}
\end{widetext}
where $\xi$ is the length of the crystal divided by the confocal parameter $b=\frac{2\pi w^2}{\lambda}$ ($w$ is the Gaussian beam radius and $\lambda$ is the wavelength), and $\mu = k_{\text{signal}} / k_{\text{idler}}$, where $k = 2\pi/\lambda$.\cite{broyer1985intracavity, canarelli1992continuous, goldberg1995difference} We plot this parameter in Fig.~\ref{fig:h_parameter} for reference.

\section{Change in Optimal PPLN oven Temperature With Power}

\begin{figure}[h!]
\includegraphics[width=8.5cm]{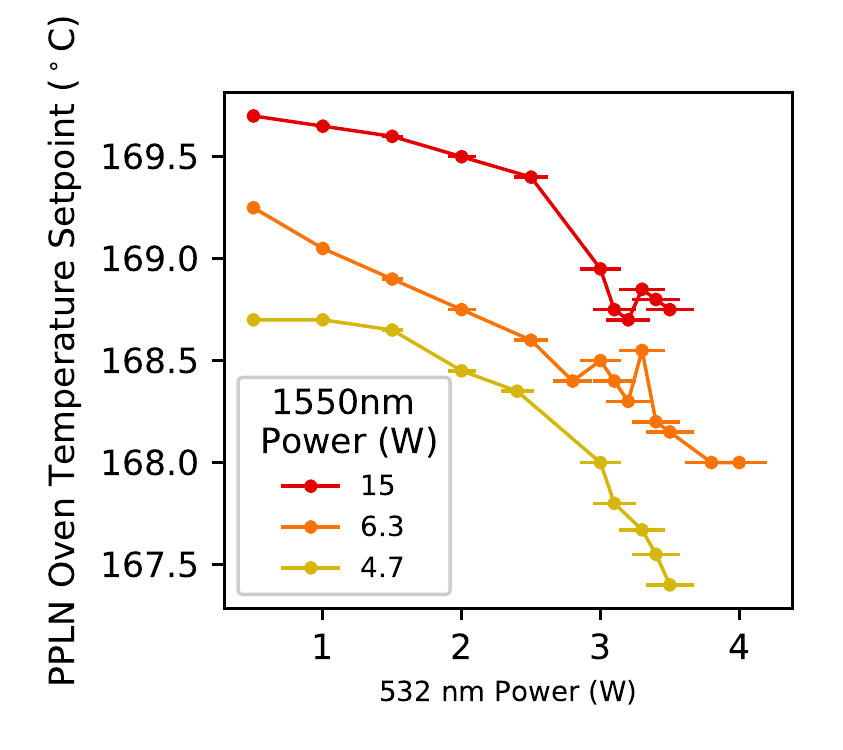}
\caption{\label{fig:Temps} Optimal temperature setpoints for data in Fig.~\ref{fig:Pow}. Vertical error bars are not presented, as we did not characterize or have access to the uncertainties in our temperature probe, which is mounted on a thermally conductive plate that is in contact with the crystal, and does not give information on the temperature gradients that are likely present. }
\end{figure}

We were not able to directly probe the temperature profile of the PPLN. However, we observed heating of the crystal through the fact its controller needed to be reoptimized for each different beam power. In Fig.~\ref{fig:Temps}, we present the power dependence of the optimal controller setpoint. For each datapoint, the temperature was swept over the optimum, with step sizes of 0.1 $^\circ$C (though smaller steps were necessary for higher power operation) and roughly one minute of settling time before each step. At the highest powers, we also observed that the optimal position for the translation stages that set the distance between the focusing lenses and the crystal would change. However, we did not optimize this distance for each power. 

\section{Residual Amplitude Noise on Source Lasers}

In order to document the source of intensity noise on our system, we note that the RIN measured on our generated light is both consistent with that on our pump laser, as well as manufacturer specifications for both the pump and the idler. At lower frequencies (<100 Hz) there is larger intensity noise on our system, perhaps due to thermal drifts of the crystal temperature, or beam alignments. However, noise at these frequencies is not limiting for laser performance, as it can be stabilized with standard techniques.\cite{wang2020reduction}

\begin{figure}[h!]
\includegraphics[width=8.5cm]{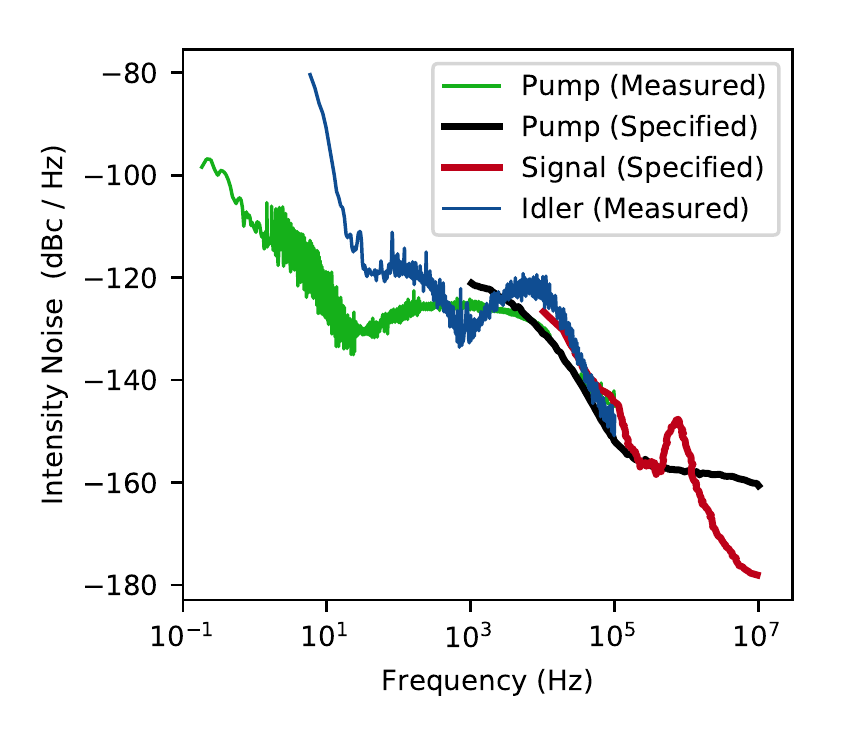}
\caption{\label{fig:RIN_Appendix} Comparison of RIN on the idler to the RIN of the pump, as well as manufacturer specified RIN values for both the pump and signal lasers. We do not measure RIN above 100 kHz, but we do not expect there to be significant intensity noise at these higher frequencies, given the specified RIN on the pump and signal lasers.}
\end{figure}

\bibliography{mybib}

\end{document}